\documentclass[12pt,preprint]{aastex}
\received{}
\accepted{}
\slugcomment{Astronomical Journal, in press}
\shorttitle{Milliarcsecond AGN of NGC 4151}
\shortauthors{Ulvestad et al.}
\journalid{}{}
\articleid{}{}
\def\H2{\ion{H}{2}}

\def\hi{\ion{H}{1}}

\begin{document}

\title{VLBA Identification of the Milliarcsecond Active Nucleus
in the Seyfert Galaxy NGC 4151}

\author{James S.~Ulvestad,\altaffilmark{1}
\altaffiltext{1}{National Radio Astronomy Observatory, 
P.O. Box O, Socorro, NM 87801; julvesta@nrao.edu}
Diane S.~Wong,\altaffilmark{2}
\altaffiltext{2}{Department of Astronomy,
University of California; dianew@astron.berkeley.edu}
Gregory B.~Taylor,\altaffilmark{3}
\altaffiltext{3}{National Radio Astronomy Observatory, 
P.O. Box O, Socorro, NM 87801; gtaylor@nrao.edu}
Jack F. Gallimore,\altaffilmark{4}
\altaffiltext{4}{Department of Physics, Bucknell University,
Lewisburg, PA 17837; jgallimo@bucknell.edu} and
Carole G.~Mundell\altaffilmark{5}
\altaffiltext{5}{Astrophysics Research Institute, Liverpool 
John Moores University, Twelve Quays House, Egerton Wharf, Birkenhead, CH41 1LD, U.K.;
cgm@astro.livjm.ac.uk}}
\setcounter{footnote}{5}

\begin{abstract}

The Seyfert galaxy NGC~4151 has been imaged at resolution better
than 0.1~pc using a VLBI array consisting 
of the VLBA and three 100m-class telescopes.
A flat-spectrum 3-mJy source with a monochromatic radio power of 
$\sim 10^{37}$~ergs~s$^{-1}$ has been detected, apparently at
the location of the active galactic nucleus (AGN) and its
central black hole.  The radio source has a minimum brightness temperature
of $2.1\times 10^8$~K and a size upper limit of 0.035~pc, about 10
times the diameter of the broad-line region and 15,000 times the diameter 
of the black hole's event horizon.  An additional flat-spectrum
component located within a parsec of the apparent nucleus is
likely to be a knot in the inner radio jet.  The presence of some
steep-spectrum radio emission within 0.1~pc of the galaxy nucleus
limits the emission measure of a possible ionized torus to
a maximum value of $10^8$~cm$^{-6}$pc.  If the hard X-ray source in
NGC~4151 is associated with the radio AGN, its radio to X-ray ratio is 
less than $10^{-5}$, putting NGC~4151 securely in the radio-quiet
class of AGNs.  The radio image reveals a 0.2-pc two-sided base
to the well-known arcsecond radio jet.  Apparent speeds of jet components
relative to the radio AGN are $<0.050c$ and
$<0.028c$ at respective nuclear distances of 0.16~pc and 6.8~pc. 
These are the lowest speed limits yet found for a Seyfert galaxy, and
indicate non-relativistic jet motions, possibly due to thermal plasma, 
on a scale only an order of magnitude larger than the broad-line region.

\end{abstract}


\section{Introduction}
\label{sec:intro}

A number of Seyfert galaxy radio sources have been imaged on 
milliarcsecond scales in recent years, using the Very Long 
Baseline Array (VLBA).  Although each galaxy
is unique, there are a number of common characteristics that have been revealed
by VLBA imaging of Seyfert and LINER (Low Ionization Nuclear Emission-line
Region) galaxies, as summarized by \citet{ulv03}: (1) Unresolved radio
cores with flat or inverted spectra often are present (Wilson et al. 1998; 
Ulvestad, Wrobel, \& Carilli 1999a; Ulvestad et al. 1999b;
Mundell et al. 2000; Falcke et al. 2000; Ulvestad \& Ho 2001; Nagar et al. 2002;
Anderson, Ulvestad, \& Ho 2004);
(2) Apparent relative motions in the few jets imaged at multiple epochs generally
are quite subluminal \citep{roy98,ulv98,ulv99b,roy00}, except during outbursts, when 
apparent speeds are seen to be as high as $0.5c$ in Mrk~348 \citep{pec03}
and near $c$ in IIIZw2 \citep{bru00}; and (3) At least some jets
change their apparent directions of motion on scales of parsecs
\citep{ulv98,ulv99b,mun03}.  In the related radio-quiet quasars, there often are
flat-spectrum cores with variability characteristics similar to the
radio-loud quasars having relativistic jets \citep{bar05}. There have been
no direct measurements of jet motions in radio-quiet quasars, though
\citet{blu03} have inferred superluminal motion in one object.

NGC~4151 was one of the original Seyfert galaxies classified in the 1940s 
(Seyfert 1943).   Its optical spectra show both broad and narrow components in
the Balmer emission lines, leading to a classification as a type 1.5 
Seyfert galaxy \citep{ost76}.  The distance of NGC~4151, derived assuming
a Hubble constant of  $H_0 = 75$~km~s$^{-1}$~Mpc$^{-1}$, is 20.3~Mpc 
\citep{tul88}.\footnote{There
are a range of distances cited in the literature for NGC~4151, typically varying
from $\sim 13$~Mpc to $\sim 20$~Mpc.  We take a distance on the high end of this
range because it has no impact on the spectral energy distribution (where all
fluxes scale together), but does give the most conservative (high) values or
limits to apparent component speeds.} This proximity and the consequent 
scale of 0.10 pc~mas$^{-1}$ make NGC 4151 one of the brightest and most 
well-studied Seyfert galaxies.  Reverberation mapping of the broad 
emission lines shows that the radius of the broad-line region 
(BLR) is $\sim 2\times 10^{-3}$~pc (Wandel, Peterson, \& Malkan 1999), 
corresponding to an angular radius of only 0.02~mas.  
Assuming that the line widths are largely virialized around a central massive
black hole, \citet{wan99} determined a 
central black hole mass of $\simeq 1.2 \times 10^{7}M_\odot$.

NGC 4151 contains a linear radio structure, or ``jet,'' 3\farcs5 (350~pc) in length at
an average position angle of $77^\circ$.  This structure has been imaged with 
progressively higher fidelity at sub-arcsecond resolution by the Very Large Array 
(VLA) \citep{wil82,joh82,ped93} and the Multi-Element Radio-Linked
Interferometer Network (MERLIN) \citep{ped93,mun95},
as well as being imaged with the European VLBI Network by \citet{har86}. 
Five main radio components in the jet (C1 through C5) were identified in
the MERLIN images, while milliarcsecond-resolution
imaging using the VLBA resolved some of the MERLIN components and 
identified at least eight separate regions of radio emission, 
A--H \citep{ulv98}.   The correspondence among the different radio components 
is shown in the top panel of Figure~\ref{fig:cgm}, reproduced
from \citet{mun03}.  The highest resolution 1.4 GHz VLBI images 
indicate that the average jet direction incorporates a number of sharp 
changes in position angle, possibly related to jet interactions with the 
circumnuclear environment \citep{mun03}.  In this paper, we discuss only the 
VLBA components D and E, corresponding to C4W and C4E in the
MERLIN scheme; no other components were detected at our high resolution.

\citet{mun95} hypothesized that the nucleus of the active galaxy is located in 
VLBA component D, based largely on the lack of \hi\ absorption toward that
radio component and the expectation that the inner part of any nuclear
torus or accretion disk would be ionized by the active galactic
nucleus (AGN).  In contrast, \citet{ulv98} suggested that VLBA component E,
located 7~pc further east, is the actual AGN, due to
the possible presence of a subcomponent with a flat radio spectrum. However, 
high-resolution VLBA imaging of NGC~4151 recently has
shown that all parts of component E exhibit \hi\ absorption, making it unlikely
that it includes the AGN, and therefore supporting the identification of
component D as the galaxy nucleus \citep{mun03}.

{\it Hubble Space
Telescope} images of NGC 4151 with Wide Field Planetary Camera 1 (WFPC1) 
and with the Faint Object Camera revealed a $3''$ narrow 
line region (NLR) with a biconical form \citep{eva93,bok95}; more details of the
conical structure and the kinematics of the NLR have been revealed in the
WFPC2 and Faint Object Spectrograph program carried out by \citet{hut98}.
Both a compact hard X-ray source and extended soft
X-ray emission have been detected at or near the nucleus \citep{elv83,wea94,yan01}.
Near-infrared variability \citep{min04} and interferometric imaging
\citep{swa03} indicate the presence of a dust torus with inner radius
of $\sim 0.05$~pc, possibly with a geometrically thin accretion disk
at smaller radius.  Placing these small components in context, 
and studying the physical properties of the core of the AGN, 
is possible only if there is a secure identification of the
radio nucleus of the galaxy.  In this paper, we report high-sensitivity
VLBI imaging of NGC~4151 that identifies the flat-spectrum radio
nucleus of the galaxy and also provides the lowest upper limit to 
date on relative parsec-scale component speeds in any Seyfert galaxy.

\section{VLBI Observations and Imaging}
\label{sec:data}

\subsection{Observations}
\label{sec:obs}

NGC 4151 was observed in dual-circular polarization at 5.0~GHz and 8.4~GHz
(6~cm and 3.6~cm wavelengths, respectively), using the ten 25m antennas 
of the VLBA \citep{nap93} together with the phased VLA (twenty-seven 25m 
antennas), for a total of 12 hours 
on 1998 March 27 (program ID BG070).  At each frequency, the total data rate
employed was 128~Mbit~s$^{-1}$, yielding a total bandwidth of 16~MHz at
each polarization, with 2 bit sampling.  In order to acquire more
sensitivity in the search for a flat-spectrum core, NGC~4151 also was observed
on 2002 May 16 (program BU022) for 10 hours, using a VLBI array consisting of the
VLBA, the phased VLA, the 100m Green Bank Telescope (GBT), and the 100m 
Effelsberg Telescope (Eb); the data rate and bandwidth were double those employed
in 1998.  In 2002, most of the observing time was spent
at 15.3~GHz (2~cm wavelength), with lesser amounts of time at 5.0 and 8.4~GHz.  
Due to the lengthy time required to reposition its subreflector in order 
to change frequency bands, the GBT observed only at 15.3~GHz.  

Because NGC 4151 is too weak to detect fringes in an atmospheric coherence time, 
phase-referencing observations \citep{bea95} were employed, using the calibration source
J1206+3941, approximately $2^\circ$ away from the galaxy.  This calibrator now
has an absolute position determined to 1~mas accuracy \citep{bea02} in the International
Celestial Reference Frame \citep{ma98}.  That position uncertainty 
dominates the absolute position errors for the NGC~4151 VLBI components, since
ionospheric errors at 5~GHz and higher should be less than 1~mas
for a $2^\circ$ phase-reference distance \citep{cha04}.  Therefore, the
absolute position error for NGC~4151 is taken to be 1~mas.  
The phase-referencing cycle time was about 12 minutes in 1998, and
2.5--4.5 minutes in 2002.  A journal of the observations, including the
total integration time on the galaxy, is provided in Table~\ref{tab:obs}.  Note
that the need for frequent calibration meant that only about 35\% of the scheduled
10 hours was spent actually integrating on the galaxy in 2002.

\subsection{Calibration}
\label{sec:calib}

All data calibration was carried out in NRAO's Astronomical Image Processing
System, AIPS \citep{gre03}.
Amplitude calibration for the VLBA antennas in 1998 and 2002 
was done by using the values for the antenna 
gains maintained by VLBA personnel as well as measurements of
system temperature made once per minute during the observations; this calibration is 
believed to have an accuracy of 5\%.  A similar method was used for calibration
of Eb during 2002; approximately once per hour,
that telescope was removed from the VLBI array in order to spend
15 minutes on pointing observations.

The 27 VLA antennas were made mutually coherent by adjusting their 
individual phases periodically during short observations of the compact 
sources J1206+3941 (in 1998) and J1146+3958 (in 2002).  The amplitude response 
of the individual VLA antennas was calibrated with respect to the standard amplitude 
calibrator J1331+3030 (3C~286), on the scale of \citet{baa77}, as modified slightly
by the most recent measurements at the VLA.  This amplitude calibration then was
transferred to the reference source J1206+3941, enabling calculation of the
complex gain, which was
interpolated in time to derive the VLA calibration for NGC~4151.

For the GBT, we initially assumed a flat gain curve and 70\% aperture
efficiency based on commissioning measurements. This initial calibration was
assessed using the amplitude check calibrators J0927+3902 and J1310+3220,
and we found it
necessary to reduce the amplitude (voltage) gain by 35\% to bring the 15 GHz
calibration in line with the VLBA scale. Similar analysis resulted in a 10\%
gain reduction for Eb, and small adjustments for several VLBA antennas,
but no modifications to the a priori calibration were necessary 
at 5 and 8~GHz.  For both 1998 and 2002, we 
estimate that the overall amplitude scale for NGC~4151 is 
uncertain ($1\sigma$) by 5\% at 5~GHz and 8.4~GHz.  The significant
amplitude corrections leave the flux-density scale at 15~GHz
somewhat more uncertain; based on ``before'' and ``after'' imaging of
the amplitude-check source and the phase-reference source, we conservatively 
estimate the $1\sigma$ uncertainty in the amplitude scale at 15~GHz to be 10\%.

Initial clock offsets for all participating VLBI telescopes (including the
phased VLA) were determined by means of phase-calibration tones and 
observations of the strong source J1642+3948 (3C~345).  Time-dependent 
instrumental (and atmospheric) delays, delay rates, and phases were 
calibrated for J1206+3941 and interpolated to the observation times of NGC 4151.  
Due to the weakness of the galaxy and the fact that useful polarization
calibration would have reduced the galaxy integration time further,
no attempt was made to calibrate the instrumental polarization;
VLA imaging \citep{ho01} indicates that the peak polarization signal on
an arcsecond scale is 0.3~mJy~beam$^{-1}$ at 4.9~GHz.

\subsection{Radio Imaging}
\label{sec:imaging}

After the best possible initial calibration had been obtained,
the calibrated data for NGC~4151 were imaged and self-calibrated using
AIPS and Difmap \citep{she97}.  Figure~\ref{fig:cgm} shows 
the VLA-only image at 15.3~GHz, on arcsecond scales, from 2002, together
with the VLBA 1.4-GHz image from \citet{mun03}.  
The VLA was in its {\bf BnA} configuration, yielding a 
resolution of 0\farcs32 by 0\farcs15.  The VLA image shows the general 
structure of the arcsecond-scale jet; note that the entire
VLBI structure discussed below, components D and E, 
is found within the source
labeled ``Nucleus'' in the VLA image.  The 15.3-GHz flux density
of this nucleus detected by the VLA is $28.1\pm 1.4$~mJy.

The VLBI imaging was carried out using a variety of weightings of the 
interferometric data; typically we used ``natural'' weighting, 
which optimizes the sensitivity, to produce the
final images.  This procedure is particularly important for the 2002 data,
where such weighting gives the greatest strength to the data provided
by the several large telescopes used along with the VLBA.  

\begin{figure}
\plotone{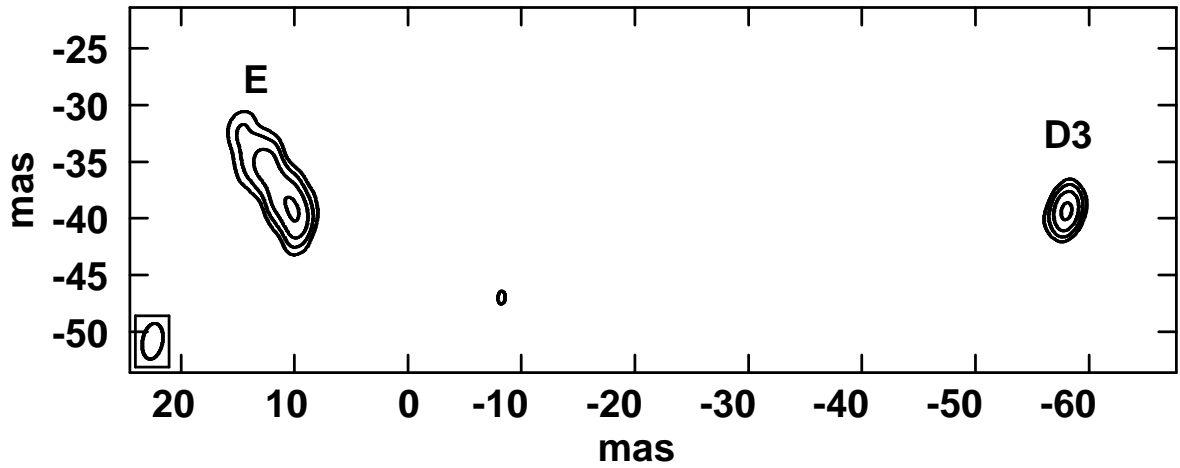}
\caption{
VLBI image of NGC~4151 at 5.0 GHz, from epoch 1998.24.
The restoring beam is $3.2\times 1.8$~mas in position angle $-10^\circ$.
Contour levels begin at 3 times the rms noise of 102~$\mu$Jy~beam$^{-1}$
and increase by factors of two.  Coordinates are given relative to 
$\alpha$=12h10m32.5831s, $\delta=39^\circ 24'21.099''$ (the original
pointing position plus an offset due to the updated position of the 
reference source J1206+3941).}
\label{fig:1998}
\end{figure}

Figure~\ref{fig:1998} shows the final 5.0~GHz VLBI image from 1998.23, where
significant emission was found at two locales separated by 
$\sim 70$~mas, or 7~pc.  These components are those designated
D3 and E by \citet{ulv98}; components D1 and D2 are not detected.  
Respective noise values at 5.0 and 8.4~GHz are 102 and 
70~$\mu$Jy~beam$^{-1}$ for the most sensitive images.
Figure~\ref{fig:2002} displays blowups of components D3 and E at 
5.0, 8.4, and 15.3~GHz, all from the data taken in 2002.37.
Because of the inclusion of the large telescopes and additional
bandwidth, the noise values are lower than those
achieved in 1998, with respective values of 48, 69, and 
45~$\mu$Jy~beam$^{-1}$ at 5.0, 8.4, and 15.3~GHz.  Observation 
and imaging parameters are summarized in Table~\ref{tab:obs}.

\section{Component Motions}
\label{sec:motions}

The nucleus of NGC~4151 is the origin of a two-sided radio 
jet that initially can be seen only within $\sim 0.2$~pc of the
AGN (see upper right panel of Figure~\ref{fig:2002}), but later is illuminated at
various locations within $\sim 200$~pc of the nucleus
\citep{mun03}.  The two-sided appearance of the jet and its
side-to-side ratio near unity indicate that it is either nearly
in the plane of the sky or is affected little by Doppler boosting.  Since we
apparently have a direct view of the broad line region in NGC~4151,
we should be viewing the jet within $\sim 45^\circ$ of its axis,
which would provide significant Doppler boosting and de-boosting of
a jet moving near $c$.  Therefore, indications just from the 
appearance of component D3 are that the local motions are considerably
below $c$ within 0.2~pc of the galaxy center.

\citet{ulv98} found an upper limit of
$0.14c$ for the relative speed between the peaks of components D3 
and E between 1984 and 1996.  The new imaging in 1998 and 2002, 
at higher sensitivity than in 1996, enables a much more accurate 
measurement to be made.  The component names in Table~\ref{tab:fluxes}
represent an attempt to identify individual components that can be 
seen at both epochs, and whose apparent motions can be measured.  
However, it should be kept in mind 
that a generally similar appearance
of images at the two epochs does not necessarily mean that
the same physical components have been seen at both epochs.  

The accuracy of the absolute positions of the various 
components in our VLBI images is limited by knowledge of the
phase-reference source position and possible atmospheric effects
over the $2^\circ$ distance between calibrator and reference
source. However, the relative positions within NGC~4151 depend only to
second order on these effects.  The dominant quantifiable
effect, instead, is simply the uncertainty in source position
set by the limited signal/noise ratio (SNR) of the data.  To zeroth order,
the error in location of a component relative to some fiducial
point in the image is equal to the full-width at half-maximum
(FWHM) of the beam divided by the SNR of a component detection.  
Self-calibration may shift the absolute position of components in
the image, but should not change the relative
positions across the image, since all data for a given VLBI
station have the same phase adjustments applied in the 
self-calibration process.  Therefore, it is appropriate to use
the SNR of the self-calibrated data in determining relative
source position errors.  

The above error analysis ignores possible systematic
errors, such as source position shifts due to changing
sensitivity to diffuse emission, frequency-dependent
effects, and uncertainties in component identification.
Apparently due to such effects, our fitting has shown
relative position shifts of up to 0.05~mas (occasionally more) 
for different data weighting in the imaging, or between 8 and 
15~GHz fitting of peaks that appeared to be the same component.   
To account for this additional error source, we add 0.05~mas
in quadrature to the SNR-determined error and the fitting errors
for each relative position measurement.

Here, note that we have been forced to add even more nomenclature, since 
component D3 splits into at least three components at 15~GHz, which
we unimaginatively call D3a through D3c (from west to east).  We
measure position shifts relative to D3b, the most likely identification
for the AGN (see Section~\ref{sec:AGN}).
In 1998.42, the 8.4-GHz separation of D3b and E1
was $68.15\pm 0.10$~mas, where we have used the beam size in
the East-West direction (roughly the direction of separation)
to compute the error.  In 2002.37, the 15-GHz separation
between the same two components was $68.33\pm 0.06$~mas.  
Similarly, we can compute the separation on a much
smaller scale, between D3b and D3c.  This changed
from $1.16\pm 0.20$~mas to $1.56\pm 0.06$~mas over the same
4-yr span; the large error in 1998.42 is caused by the
difficulty in separating D3b and D3c at 8.4 GHz.

On the face of it, the above analysis indicates possible 
detections of motion at slightly less than a $2\sigma$ level 
in components located at 0.16~pc and 6.8~pc from the AGN.  
However, since the possible systematic and frequency-dependent
effects are of somewhat uncertain magnitude,
we would need at least a $3\sigma$ 
result to be confident of a detection of motion in the radio 
components.  Therefore, we quote $3\sigma$ upper limits for the 
motion relative to D3b in 4.14~yr; these upper limits are 0.63~mas
for component D3c and 0.35~mas for component E1, corresponding 
to respective velocity upper limits of $0.050c$ and $0.028c$.
Assuming that the component identification across two epochs 
is correct, this is the lowest speed limit yet measured for
any Seyfert galaxy.  Of course, we must cite the usual {\it caveat}
that we cannot distinguish motion (or lack thereof) of the actual
radio-emitting material from motion (or lack thereof) of some
structure such as a shock, which may stay near one position even
while material flows through it.  Still, the very low 
limit of $~\sim$15,000~km~s$^{-1}$ within 0.16~pc of the AGN 
is not much larger than the gas velocities within the broad line 
region, and indicates that the jet in NGC~4151 may be dominated
by thermal plasma in its innermost regions.

\section{Identification of the Active Nucleus}
\label{sec:AGN}

At radio wavelengths, AGNs are most easily identified by the
presence of a compact, flat- or inverted-spectrum core.  \citet{ulv98}
inferred that such a core was most likely to be present in
component E, in order to account for the slightly flattened
radio spectrum attributed to component E.  A principal aim of
the present observations was to achieve considerably higher
sensitivity to check for the presence of a weak core
with a flat radio spectrum.

\begin{figure}
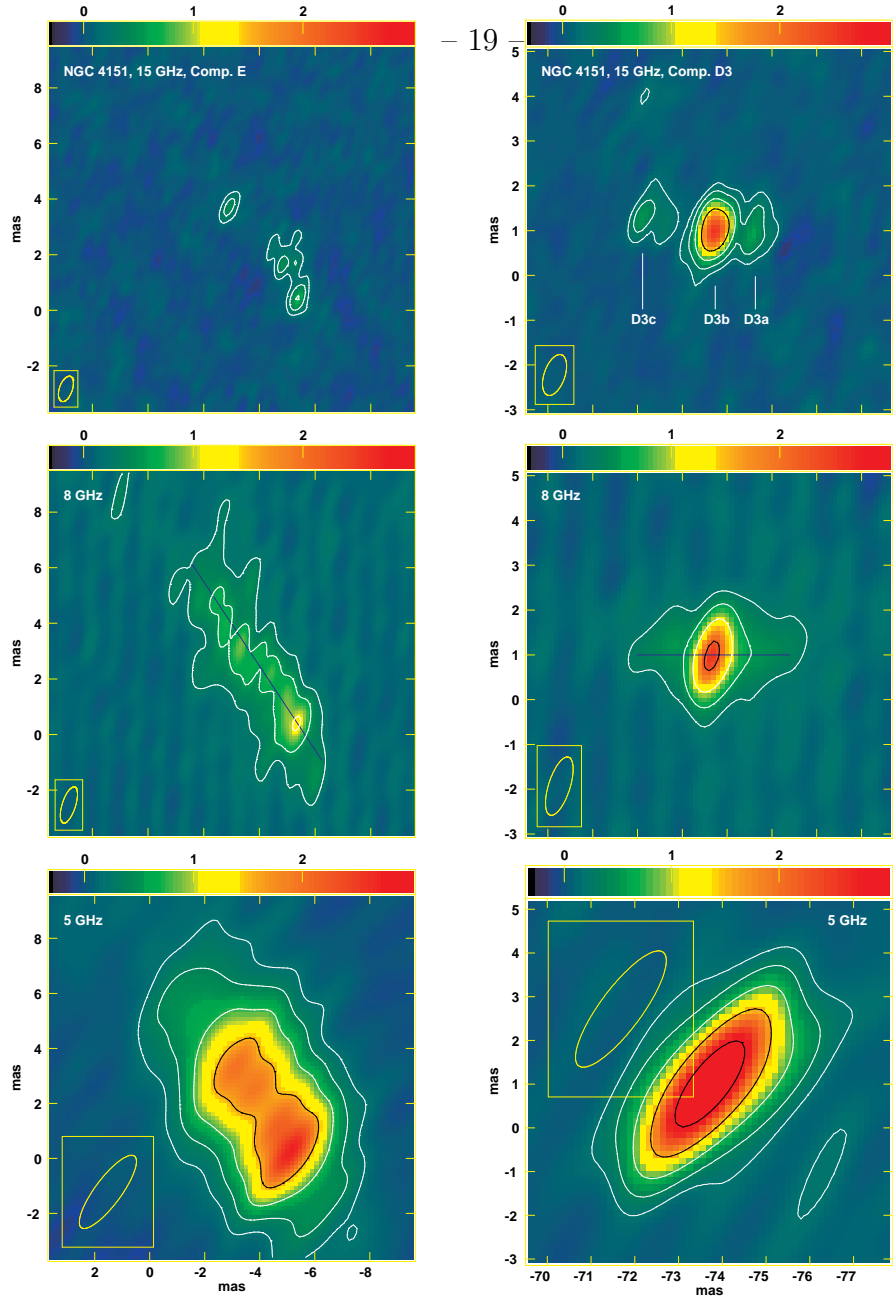

\vspace{16cm}
\includegraphics{./f3a.eps}
\includegraphics{./f3b.eps}
\includegraphics{./f3c.eps}
\includegraphics{./f3d.eps}
\includegraphics{./f3e.eps}
\includegraphics{./f3f.eps}
\caption{
From top to bottom, 15.3, 8.4, and 5.0 GHz VLBI images of 
NGC~4151 are shown at epoch 2002.37.  Contour levels begin at 4 times 
the rms noises given in Table~\ref{tab:obs} and increase by factors of two.
Beam sizes are given in Table~\ref{tab:obs} and shown by the boxed ellipses in
lower or upper left-hand corners.  Component E is shown in the left 
panels, with D3 in the right panels, at a somewhat different scale.  
The lines drawn along the major axes of
the 8.4 GHz images indicate the positions of the slices
shown in Figures~\ref{fig:sliceE} and \ref{fig:sliceD}.  
All coordinates are given in 
milliarcseconds offset from the original pointing position of
$\alpha$=12h10m32.5822s, $\delta=39^\circ 24'21.059''$.  All images 
have the same flux-density color scale, shown by the bar at the
top of each image.}
\label{fig:2002}
\end{figure}

Inspection of Figure~\ref{fig:2002} immediately shows that 
component E contains no compact emission at 15~GHz 
whose flux density is above 1~mJy.  Since the peak flux
densities at lower frequencies are much higher, there is
no obvious flat-spectrum component that might be identified
as the galaxy nucleus.  However, component D3
contains an unresolved radio source with a flux density
of approximately 3~mJy at 15~GHz, at the AGN location 
inferred by \citet{mun95,mun03}.

Of course, a visual impression is inadequate without a
quantitative test for the presence of a flat-spectrum
nucleus.  The search for such a nucleus is complicated
by the presence of several components in both D3
and E, combined with the quite different resolutions
of the observations at frequencies varying by a factor of
three.  In addition, in component E at the low
frequencies, the compact peaks clearly are embedded in
smoother underlying emission that makes it challenging
to isolate compact emission regions.

We measured unresolved component flux densities by
fitting Gaussian components to each peak in the 
images made with various data weightings and resolutions.
Contamination from the underlying smooth emission was minimized
by constraining the compact components to be unresolved.
The fitted flux densities for images made 
with different data weighting are consistent with one
another, so we use an average from the different images for 
each component.  The combination of fitting and amplitude
errors for individual components is
estimated to be 7\% at 5 and 8~GHz, and 12\% at 15~GHz. 
These errors are larger than the amplitude scale uncertainty
discussed in Section~\ref{sec:calib} because of 
increased confusion by the smooth emission 
underlying some compact components, as estimated by using 
component fits in images made with different data weighting.  
The increases of the component amplitude errors are fractionally
larger at the lower frequencies, where the resolution is
poorer and the underlying smooth flux is brighter.
The final uncertainties in component flux densities were determined
by taking the above errors from the amplitude scale and component
fitting, then adding them in quadrature to the rms noise.  

In order to test quantitatively for the presence of a 
flat-spectrum nucleus, we have tapered the weighting of the 15-GHz 
data in the visibility plane in order to achieve resolution similar
to that at 8~GHz.
The resulting images (not shown, since they are similar to
those in Figure~\ref{fig:2002}) then were analyzed in two ways.  
First, we fit independent Gaussians to each significant component;
results of those fits, including the relative source positions 
good to 0.01~mas, are presented in Table~\ref{tab:fluxes}.  
In component E, the radio emission is fairly smooth at 5 and 8 GHz,
and shows a few local maxima at 15 GHz.  Fitting individual compact 
components in component E was largely an exercise in intuition
at the lower frequencies. Therefore, we have fitted the strongest 
emission component in E, calling it E1, but otherwise have not 
split component E into multiple subcomponents.  
E1 has a steep radio spectrum, with a two-point spectral index
between 8 and 15 GHz of $\alpha_{8,15}=-1.27\pm 0.16$
(defining $S_\nu\propto \nu^{+\alpha}$). 
In contrast, D3b, the most powerful component of D3, has 
$\alpha_{8,15}=-0.11\pm 0.23$.
Component D3c has $\alpha_{8,15}=-0.48\pm 0.30$, consistent
with either a flat or a steep spectrum, while D3a is not clearly
separable from D3b at 8 GHz, and therefore cannot be fitted.

\begin{figure}
\plotone{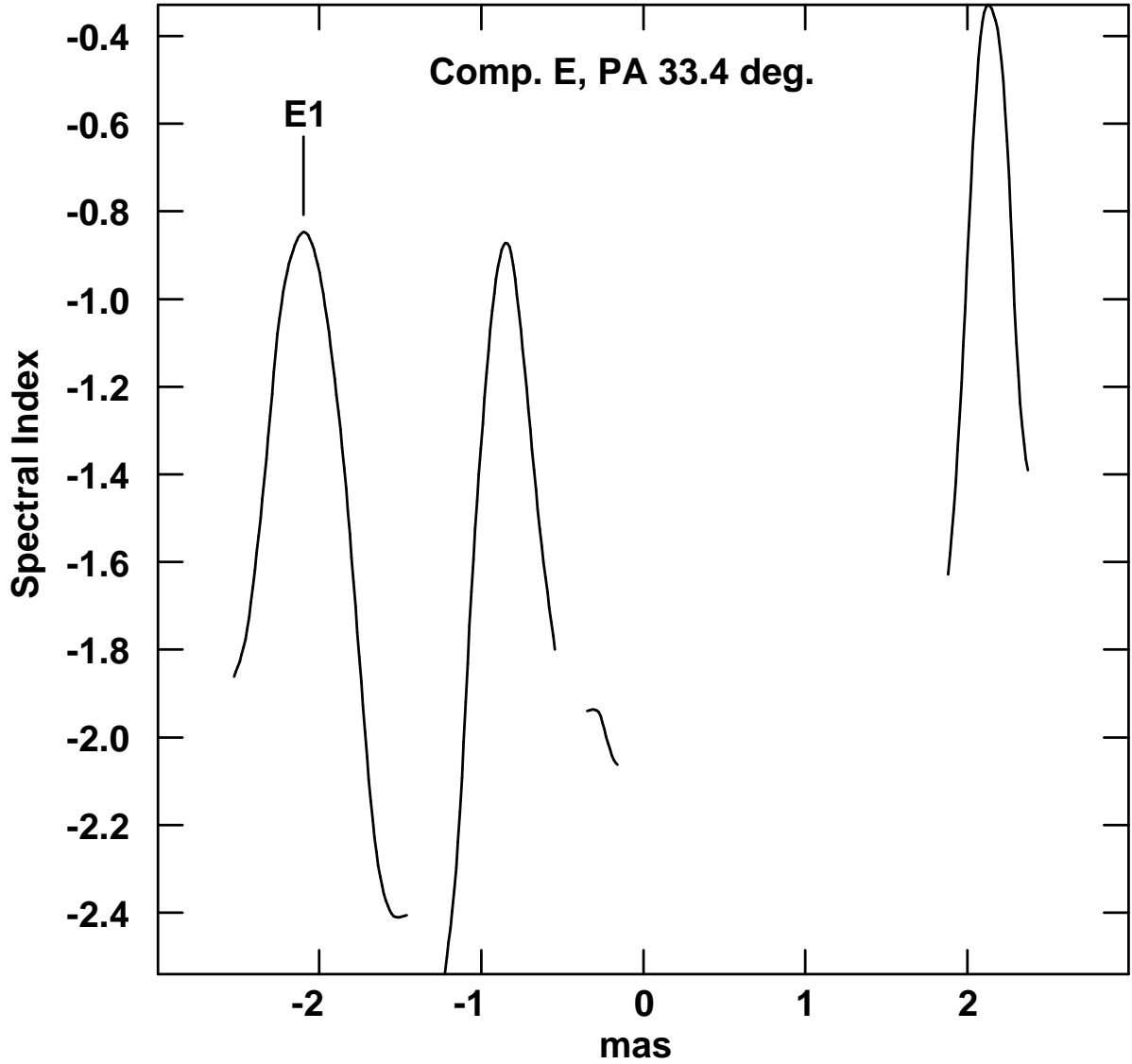}
\caption{
Two-point spectral index along the major axis of component 
E between 8.4 and 15.3 GHz at 2002.37, using the 15.3~GHz image
smoothed to the 8.4-GHz resolution.  
Distances are labeled in milliarcseconds relative
to the center of the slice, and the position of E1 is
indicated.  The spectral index map has been blanked 
where the flux density is below the $5\sigma$ detection
threshold in either input image.  The 
slice goes from SW to NE, in position angle $33.4^\circ$.}
\label{fig:sliceE}
\end{figure}

The second method of analysis was to produce an image of the
spectral index between 8 and 15 GHz. This was done by 
aligning the 8 and 15 GHz images on the strongest peak of D3, 
then computing the spectral index at each individual pixel and generating
a new spectral-index image; this image was blanked at all points
where either of the input images had a flux density less than
five times the noise level.  The spectral indices then were
evaluated along major axis slices through components D3 and E.
Figure~\ref{fig:sliceE} shows the spectral index along 
the major axis slice of Component E.  There clearly is no
flat- or inverted-spectrum component here, so it is unlikely that 
Component E contains the nucleus of NGC~4151.

\begin{figure}
\plotone{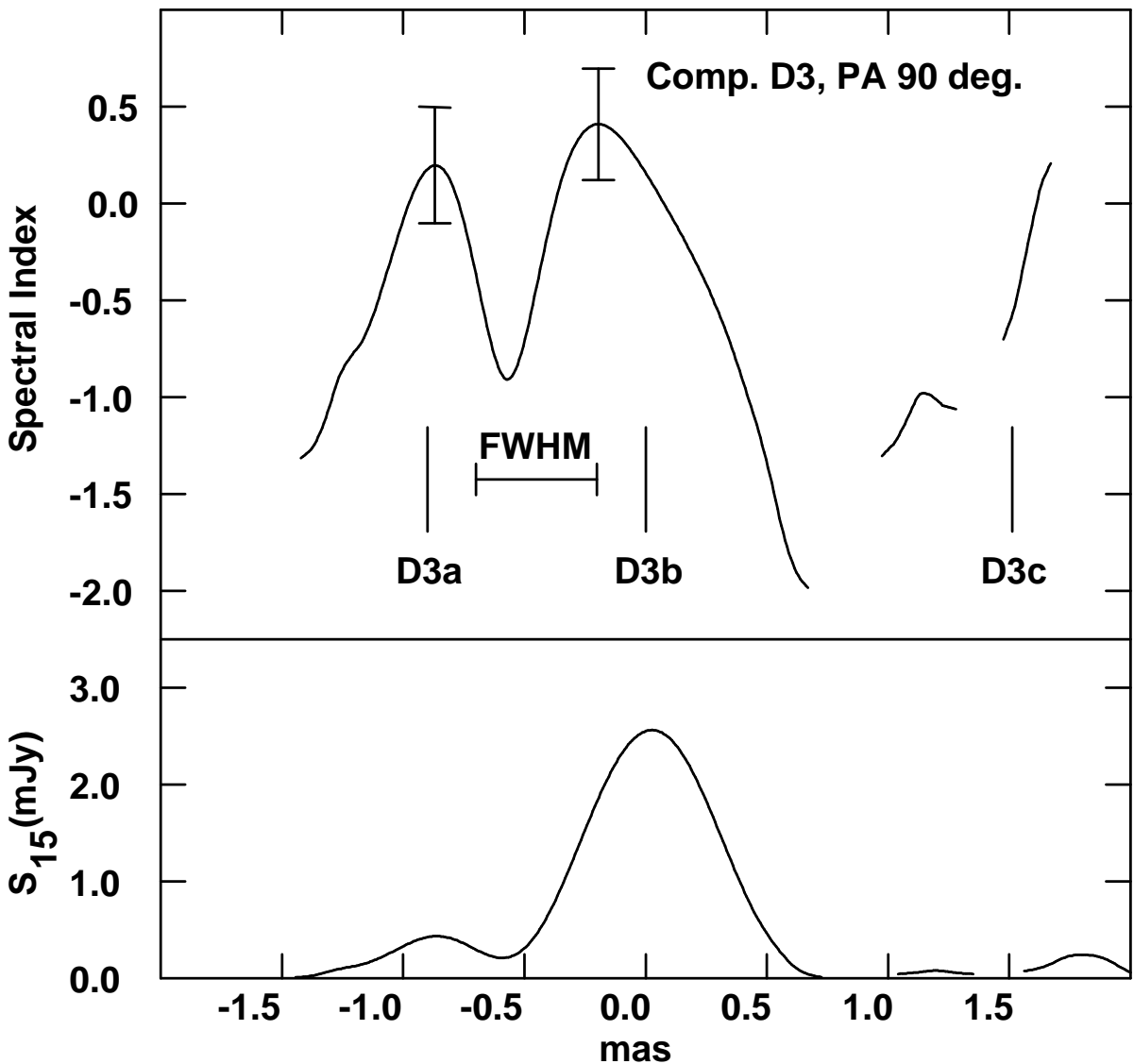}
\caption{
Slices of two-point spectral index (top panel) and 
15-GHz flux density (bottom panel) 
along the major axis of component D3, in position angle $90.0^\circ$, 
using the same images as in Figure~\ref{fig:sliceE}.
Positions of the three sub-components of D3 are shown, as are
the representative ($1\sigma$) spectral-index errors and the
full width at half maximum (FWHM) of the image resolution along
the slice.  Note that the entire slice is only a few resolution 
elements in length so small features should be interpreted with 
extreme caution.}
\label{fig:sliceD}
\end{figure}

Figure~\ref{fig:sliceD} shows the spectral index (top panel)
and the 15 GHz total intensity (bottom panel) along the major
axis (in the direction from West to East) of Component D3;
the locations of the three sub-components in
D3 are indicated in the figure.  It appears that there is
flat-spectrum emission near all three radio components.
We note, however that the major axis slice does not intersect
the peak of D3c, which is considerably north of the
major axis defined by the two stronger components.  The
apparent flat spectrum near D3c appears to be an artifact of
the fact that the slice actually intersects the edge of the
radio component, where the SNR is quite
low and the spectrum probably is affected by component 
registration errors that are a small fraction of a beam
in size.

We then must assess the spectra of D3a and D3b from the 
spectral slices.  The spectral-index image was
produced by aligning the 8 GHz and 15 GHz peaks, in D3b,
to an accuracy of about 1 microarcsecond.  Assuming that
this alignment is correct, the spectral index at the
exact location of D3b can be taken from the spectral-index
slice, and is found to be $\alpha_{8,15}=+0.16\pm 0.23$, 
consistent with the value of $-0.11\pm 0.23$ that was found
from the gaussian fitting.  At D3a, we find a similar
spectral index of $\alpha_{8,15}=+0.20\pm 0.24$.  This
cannot be compared with the value from a gaussian fit
because the 8-GHz image has no distinct peak at the location
of D3a.  We note that taking single values from a spectral
index slice is a perilous endeavor, because this process
amounts to assuming both infinite resolution and perfect
alignment of the two input images.  The beam size
along the major axis of D3 is 0.5~mas, as plotted in 
Figure~\ref{fig:sliceD}, and the absolute alignment of the
8 GHz and 15 GHz images probably is known to no better than
0.1--0.2~mas.  Therefore, the best conclusions that can be 
drawn from the spectral-index slices are that D3a and D3b
represent the location of two flat-spectrum components separated 
by less than twice the image resolution, and that there appears to
be weak steep-spectrum emission between them.

A similar analysis to that described above might be carried out using 
the data at 5~GHz as well.  However, there is insufficient resolution 
of D3 to distinguish the separate components, which causes any possible
nuclear source to be blended with surrounding emission.  If we simply
use the values in Table~2, we find $\alpha_{5,15}=-0.26\pm 0.13$ for
component D3b.
Since the 5-GHz flux density includes some confusing 
flux from component D3c, the true spectral index of
D3b is probably less steep (more positive, or less negative)
than the computed value.  

\begin{figure}
\vspace{14cm}
\includegraphics{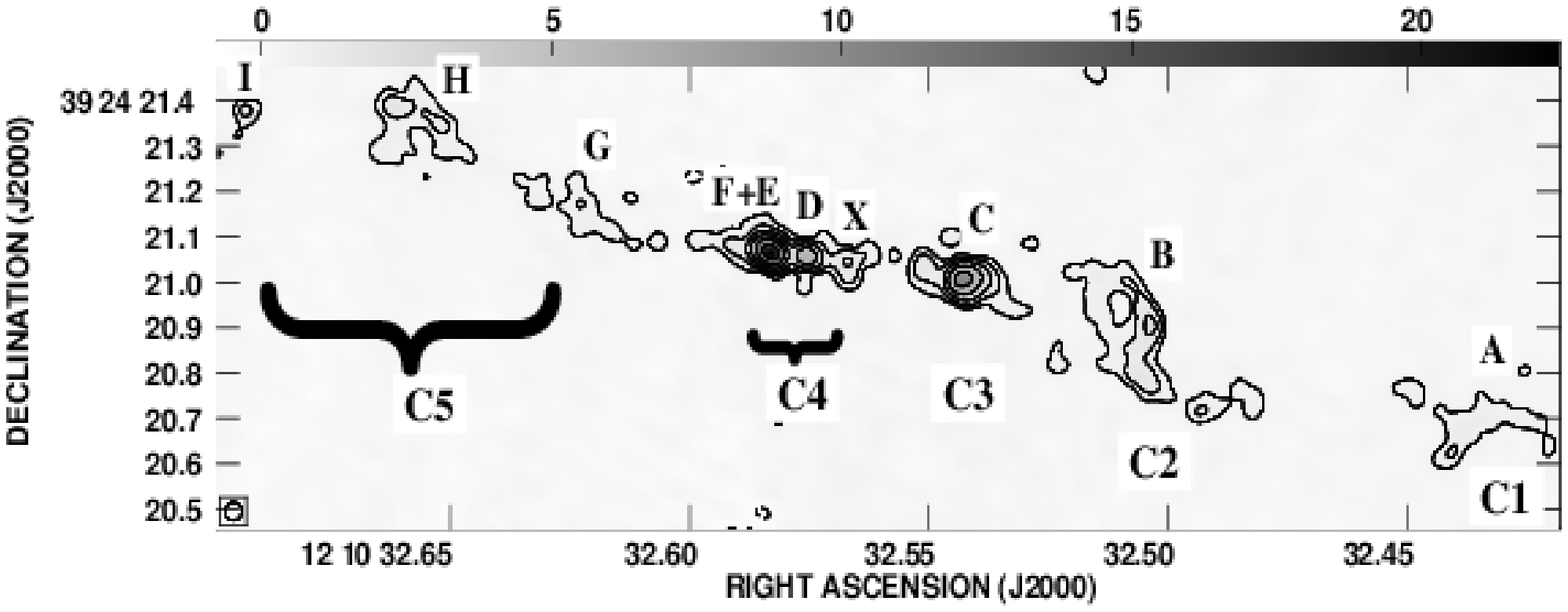}
\includegraphics{./f1b.eps}
\caption{
{\it Top:} VLBA Image of NGC 4151 jet, from \citet{mun03}, showing the nomenclature
for the radio components.  {\it Bottom:} 15 GHz VLA image of NGC~4151 at 
epoch 2002.37 from the current observations.  Components, X, D, E, and F
from the upper panel all are included in the component marked ``Nucleus'' in
the bottom panel.  The restoring beam for the VLA image 
is 0\farcs32 by 0\farcs15 in position angle 84\arcdeg.
Contour levels begin at 3 times the rms noise of 141~$\mu$Jy~beam$^{-1}$
and increase by factors of two. The
scale of the VLA image differs slightly from the VLBA image, due to the
much larger beam size.}
\label{fig:cgm}
\end{figure}

The spectral data strongly support the inference of 
\citet{mun95,mun03} that component D3 contains the true AGN in 
NGC~4151. D3b is the strongest flat-spectrum component, by 
about a factor of four.  In addition, the two-sidedness of the
arcsecond-scale jet (see Figure~\ref{fig:cgm}), as well
as the predominance of two-sided jets in low-power radio
sources \citep{gio05} make it likely that the smallest
scale jet should be two-sided rather than one-sided.
These factors imply that D3b is the most likely AGN identification,
and such an identification will be assumed below.  However,
the reader should bear in mind that it is conceivable that
the AGN might be located at D3a.  In that case, the nuclear radio-source
size for self-absorption (see below) would be smaller by about
a factor of two, the innermost jet would be one-sided, and the
nuclear radio/X-ray ratio would be smaller by a factor of 
about four.  \citet{mun03} suggested that
NGC~4151 might be similar to the compact symmetric objects
seen in more distant radio galaxies, but the apparent lack
of any flat-spectrum component between D3a and D3b makes it
highly unlikely that they represent jet components on opposite
sides of an undetected radio nucleus.

Absolute J2000 coordinates of D3b, are $\alpha$=12h10m32.5758s, 
$\delta=39^\circ 24'21.060''$, accurate to 1~mas.
This radio nucleus should be at the apex of the ``cone'' of
narrow-line clouds imaged with HST \citep{eva93,bok95,hut98},
and presumably at the center of the broad-line region 
\citep{wan99} as well as the center of the accretion disk 
and dust torus detected in
the near infrared \citep{swa03,min04}.  Since the Gaussian fit of D3b 
indicates that it is unresolved, we take an upper limit of 
half the highest-resolution beam size in each dimension, yielding a
brightness temperature of $T_{\rm B} > 2.1\times 10^8$~K.
This is consistent with, but does not require, the possibility 
that synchrotron self-absorption could produce the flat radio
spectrum.  The angular upper limit of 0.35~mas (averaging the
major- and minor-axis limits) for the radio source size 
corresponds to a linear diameter of less than 0.035~pc.
This is only $\sim 10$ times the diameter of the BLR, 
but about 15,000 times the diameter of the event 
horizon of the supermassive
black hole.  If we assume that the flat spectrum is due to
synchrotron self-absorption, and that this requires
$T_{\rm B}\gtrsim 10^{10}$~K, the upper limit to the radio
source size would be very close to the diameter of the
BLR, or $\sim 4\times 10^{-3}$~pc.  This limit 
of about 2000 gravitational radii is comparable to 
the limits for several low-luminosity AGNs in galaxies found 
at similar distances \citep{ulv01,and04}.

The apparent flat-spectrum component D3a, which appears
relatively prominent at 15 GHz and not obvious at 8.4~GHz,
has a lower brightness temperature limit of 
$T_{\rm B} > 5\times 10^7$~K.  This component could well
represent a shock in the inner jet, less than 0.1~pc from
the AGN.  One might speculate that D3a could be a location 
where the jet collides with material in the inner narrow-line
region, as may be seen on larger scales \citep{mun03}.  However,
the inability of optical emission-line imaging to probe a scale
similar to that seen by VLBI makes it difficult to test this
possibility.

An alternative hypothesis for the flat-spectrum radio components
would be that their spectral shapes are caused by free-free
absorption by ionized gas in the inner torus, lying in front of
synchrotron-emitting plasma.  Since D3a and D3b have fairly flat
spectra between 8 and 15 GHz, the presence of free-free absorption
would imply an optical depth $\tau_{\rm ff}\approx 1$ at a frequency of 
$\sim 10$~GHz; for a gas temperature near 8000~K, this implies
an emission measure of $n_e^2 dL\approx 3\times 10^8$~cm$^{-6}$pc.
If all of component D3 were enveloped in an ionized torus of
radius $\sim 3$~pc \citep{mun03}, then the average density in that torus 
(for a filling factor of unity) would be $\sim 10^4$~cm$^{-3}$,
somewhat below the value inferred by \citet{mun03}.  However,
it seems apparent that there is some steep-spectrum emission
in D3 (e.g., between D3a and D3b, as shown in Figure~\ref{fig:sliceD}).
This would imply a very patchy torus, and would make one wonder
why the densest ionized gas ``coincidentally'' lies in front of
the radio components.  Thus, the more likely scenario for free-free
absorption, if it caused the flat spectra, might be an ionized ``skin'' 
around the radio components.  This thermal gas
would have a path length shorter than 0.1~pc, and therefore
an ionized density $n_e \gtrsim 10^5$~cm$^{-3}$.  In fact, we consider
the most likely inference to be that we have detected synchrotron
self-absorption and no free-free absorption.  Then,
the presence of some steep-spectrum radio emission at 8~GHz
simply limits the overall emission measure of a quasi-uniform
ionized medium to $\lesssim 10^8$~cm$^{-6}$pc.

We note that the flux density measured for component
D3 at 15.3~GHz is above the upper limit previously reported
by \citet{ulv98}.  The difference might result from source variability by
a factor of two, certainly a viable possibility given the strong 
variability of the hard and medium-energy X-rays from 
the galaxy nucleus \citep{per81,wea94}.  However, the 1996.43 VLBI 
phase-referencing observations \citep{ulv98} also could have
suffered from substantial loss of atmospheric coherence.  In 2002.37, 
there was significant coherence loss in the initial phase referencing,
which was recovered by the use of the large telescopes both for 
the initial detection of component D3 and for the subsequent 
self-calibration.  Therefore, the present data seem to be consistent
with the previous upper limit of 2.2~mJy~beam$^{-1}$ at 15 GHz,
once coherence losses are taken into account.

The total flux density in VLBI-detected components is only $6.4\pm 0.8$~mJy
at 15~GHz; this number was derived by adding the individual fits in 
component D to the integrated flux density for the entire emission
region in E.  This is only $\sim$20\%--25\% of the total core flux 
density seen by the VLA (see Figure~\ref{fig:cgm}), with only 10\% of 
the total VLA flux density found in the putative AGN.  Therefore, much 
of the unresolved VLA 15-GHz emission must be in jet components that are
resolved out by the VLBA.  This result points out the peril of using
even unresolved VLA flux densities in modeling the emission processes
from AGNs, since the spectral energy distribution of the actual
milliarcsecond AGN, and the radio/X-ray ratio, would be in error by an 
order of magnitude if the unresolved VLA flux were used.

The 2--9~keV X-ray flux detected by the {\it Chandra} X-ray satellite
is $4.8\times 10^{-11}$~ergs~cm$^{-2}$~s$^{-1}$ (Yang et al. 2001),
corresponding to a luminosity of $2.3\times 10^{42}$~ergs~s$^{-1}$.
Using a factor of $\sim 7$ to convert from X-ray to bolometric luminosity 
\citep{ho99,ho00}, this implies that the nucleus of NGC~4151 is 
radiating at about 1\% of its Eddington luminosity of 
$1.6\times 10^{45}$~ergs~s$^{-1}$ for the $1.2\times 10^7M_\odot$ black 
hole inferred by \citet{wan99}.  Component D3b has a flux density of
$\sim 4$~mJy at 5~GHz, and a fairly flat spectrum,
implying a monochromatic 5-GHz luminosity of 
$\nu S_\nu\sim 1\times 10^{37}$~ergs~s$^{-1}$.  Given that the
quoted X-ray luminosity was near the minimum for a source that varies by 
about a factor of 10 (Weaver et al. 1994; Yang et al. 2001), 
the ratio of radio to hard X-ray luminosities is given by 
$R_X = \log[\nu S({\rm 5 GHz})/F_X({\rm 2-10 keV})]\approx-5.5$ to $-6.5$,
roughly two orders of magnitude lower than the similar ratio for
several low-luminosity active galaxies \citep{ulv01,ter03}.
$R_X$ is clearly much less than $-4.7$, the
dividing line below which an AGN such as NGC~4151 would be considered
radio-quiet \citep{ter03}.\footnote{The reader should keep in mind that the
X-ray and radio strengths were not measured at the same epoch, and that
there is no conclusive observational evidence for or against the 
possibility that the weak radio core might vary by factors of 2--3.
However, an increase by a factor of three in the radio core still 
would leave it in the radio-quiet regime as defined by the radio/X-ray ratio.}

\section{Conclusions}
\label{sec:conc}

We have imaged the core of the Seyfert galaxy NGC~4151 at two
epochs and multiple frequencies with a VLBI array consisting of 
the VLBA and several 100m-class radio telescopes.  The images
reveal a compact flat-spectrum radio component having 
$T_{\rm B} > 2.1\times 10^8$~K, which we identify as
the actual location of the active galactic nucleus and its
12-million solar mass black hole; this component is at the
location inferred by Mundell et al. (1995, 2003).  The radio flux
density of $\sim$3--4~mJy corresponds to a monochromatic
power of $\sim 10^{37}$~ergs~s$^{-1}$ at 5--15~GHz; comparison
to the X-ray luminosity indicates that NGC~4151 is a radio-quiet
object, in contrast to a number of other low luminosity active galaxies.
A weak, two-sided beginning to the larger scale radio jet is
seen to exist well within the inner parsec of the AGN.  Upper
limits to the component speeds relative to the apparent core are
$0.050c$ and $0.028c$ at respective distances of 0.16~pc and
6.8~pc from the AGN, implying that the NGC~4151 jet is non-relativistic,
and dominated by thermal plasma,
all the way down to near the broad-line region.  This is consistent
with the approximately symmetric radio morphology about the apparent center of
activity.

\acknowledgments

The National Radio Astronomy Observatory is a facility of the 
National Science Foundation operated under cooperative agreement by Associated 
Universities, Inc.  We thank the staffs of the VLA, VLBA, GBT, and Effelsberg
telescopes that made these observations possible; we are especially grateful
to Frank Ghigo for his efforts in making these VLBI
observations a success very soon after the 15~GHz capability first became
available at the GBT.
DSW acknowledges support from the NRAO summer research program and from the
National Science and Engineering Research Council of Canada.  CGM acknowledges
financial support from the Royal Society.  We thank the anonymous referee
for very useful suggestions about the analysis of the nuclear radio spectrum
and other issues.

\clearpage

\clearpage

\begin{deluxetable}{lccccc}
\tabletypesize{\scriptsize}
\tablecolumns{6}
\tablewidth{0pc}
\tablecaption{VLBI Observations of NGC~4151}
\tablehead{
\colhead{(1)}& \colhead{(2)}& \colhead{(3)}& \colhead{(4)}&
\colhead{(5)}& \colhead{(6)} \\
\colhead{Date}&\colhead{Frequency}&\colhead{Telescopes}&\colhead{Integration}&
\colhead{Resolution}&\colhead{rms noise} \\
\colhead{}&\colhead{(GHz)}&\colhead{}&\colhead{(min)}&\colhead{(mas)}&
\colhead{($\mu$Jy beam$^{-1}$) } }
\startdata
1998 Mar 27 & 4.991&VLBA,VLA&62 &$3.19\times 1.76$, pa $-10^\circ$ 
&102 \\
&8.421&VLBA,VLA&316 &$1.62\times 1.22$, pa $-1^\circ$ &70 \\
2002 May 16&4.991&VLBA,VLA\tablenotemark{a}, Eb\tablenotemark{b}&
30& $3.24 \times 0.95$, pa $-37^\circ$ &48 \\
&8.421&VLBA,VLA\tablenotemark{a}, Eb\tablenotemark{b}&
37 &$1.36\times 0.48$, pa $-18^\circ$ &69 \\
&15.365&VLBA,VLA\tablenotemark{a}, Eb\tablenotemark{b}, GBT\tablenotemark{c}&
142& $0.95\times 0.46$, pa $-19^\circ$ &42 \\
\enddata
\tablenotetext{a}{The VLA had a reduction of $\sim 15$\% in
the effective time on source in order to phase the array and to observe a
flux-density calibrator.}
\tablenotetext{b}{Eb had a reduction of $\sim 50$\% in the
effective time on source for pointing checks and because the
galaxy set at Eb part-way through the observing session.}
\tablenotetext{c}{The GBT observed only at 15.3~GHz in 2002.}
\label{tab:obs}
\end{deluxetable}

\clearpage

\begin{deluxetable}{lccccc}
\tablecolumns{6}
\tablewidth{0pc}
\tablecaption{Radio Component Flux Densities\tablenotemark{a}}
\tablehead{
\colhead{(1)}& \colhead{(2)}& \colhead{(3)}& \colhead{(4)}& \colhead{(5)}& \colhead{(6)}  \\
\colhead{Component}&\colhead{$\Delta\alpha$}&\colhead{$\Delta\delta$}&
\colhead{$S_5$}&\colhead{$S_8$}&\colhead{$S_{15}$} \\
\colhead{}&\colhead{(mas)}&\colhead{(mas)}&\colhead{(mJy)}&
\colhead{(mJy)}&\colhead{(mJy)} }
\startdata
\multicolumn{6}{c}{\underbar{1998}} \\
D3a&\nodata&\nodata&\nodata&\nodata&\nodata \\
D3b&0.00&0.00&$2.9\pm 0.2$&$2.7\pm 0.2$&\nodata \\
D3c&$+1.16\pm 0.20$&$+0.02\pm 0.20$&\nodata&$0.7\pm 0.1$&\nodata \\
E1&$+68.15\pm 0.10$&$-0.19\pm 0.10$&\nodata&$1.7\pm 0.1$&\nodata \\
\multicolumn{6}{c}{\underbar{2002}} \\
D3a&$-0.90\pm 0.06$&$0.01\pm 0.12$&\nodata&\nodata&$0.7\pm 0.1$ \\
D3b&0.00&0.00&$4.0\pm 0.3$&$3.2\pm 0.2$&$3.0\pm 0.4$ \\
D3c&$+1.51\pm 0.06$&$+0.37\pm 0.10$&\nodata&$0.8\pm 0.1$&$0.5\pm 0.1$ \\
E1&$+68.33\pm 0.06$&$-0.52\pm 0.09$&$3.0\pm 0.2$&$1.5\pm 0.1$&$0.7\pm 0.1$ \\
\enddata
\tablenotetext{a}{All coordinates are given relative to component
D3b, which has a J2000 position (good to 1~mas) of
$\alpha$=12h10m32.5758s, $\delta=39^\circ 24'21.060''$.}
\label{tab:fluxes}
\end{deluxetable}

\end{document}